\documentclass[a4paper]{IEEEtran}
\usepackage[boxruled]{algorithm2e}
\usepackage{enumerate}
\usepackage{cite}
\usepackage{graphicx}
\usepackage{psfrag}
\usepackage{subfigure}
\usepackage{url}
\usepackage{amsmath}
\usepackage{amsthm}
\usepackage{array}
\usepackage{algorithm2e}
\usepackage{amssymb}
\usepackage{amsfonts}
\usepackage{float}
\newtheorem{proposition}{Proposition}
\newtheorem*{conjecture*}{Conjecture}
\newtheorem{lemma}{Lemma}

\newtheorem{remark}{Remark}
\newtheorem{definition}{Definition}
\newtheorem{theorem}{Theorem}
\newtheorem{example}{Example}

\title{Uniprior Index Coding}
\begin{document}
\author{
}
\date{\today}
\author{
\IEEEauthorblockN{ Vijaya Kumar Mareedu and Prasad Krishnan\\}
\IEEEauthorblockA{Signal Processing and Communications Research Centre,\\
International Institute of Information Technology, Hyderabad.\\
Email: \{vijaya.kumar@research.,prasad.krishnan\}@iiit.ac.in\\}
}
\maketitle
\thispagestyle{empty}	
\pagestyle{empty}
\begin{abstract}
The index coding problem is a problem of efficient broadcasting with side-information. We look at the uniprior index coding problem, in which the receivers have disjoint side-information symbols and arbitrary demand sets. Previous work has addressed single uniprior index coding, in which each receiver has a single unique side-information symbol. Modeling the uniprior index coding problem as a \textit{supergraph}, we focus on a class of uniprior problems defined on \textit{generalized cycle} supergraphs. For such problems, we prove upper and lower bounds on the optimal broadcast rate. Using a connection with Eulerian directed graphs, we also show that the upper and lower bounds are equal for a subclass of uniprior problems. We show the NP-hardness of finding the lower bound for uniprior problems on generalized cycles. Finally, we look at a simple extension of the generalized cycle uniprior class for which we give bounds on the optimal rate and show an explicit scheme which achieves the upper bound.
\end{abstract}
\section{Introduction}
The index coding (IC) problem \cite{BiK} consists of a source generating messages connected to a set of receivers via a broadcast channel, each of which have some demands and possess some prior side-information. An optimal index code for a given configuration of demands and side-information is an encoding of the messages with the least broadcast rate, such that the demands are decodable. Different classes of the IC problem were studied based on the configuration of the side-information symbols and the demands. In \textit{unicast IC} \cite{BBJK}, the demand sets of the receivers are disjoint. The unicast IC problem can always be studied as \textit{single unicast}, where each receiver demands a unique message. In \textit{uniprior IC} \cite{OnH}, the side-information sets at the receivers are disjoint. The general multiprior/multicast index coding problem has been studied in \cite{TDN}. All such IC problems have been extensively studied using approaches from graph theory (see for example \cite{BBJK,DSC,BKL2,TDN,OnH}). While the optimal broadcast rate for index codes has been characterised precisely for only a relatively small classes of problems (single uniprior is one of them), for most classes finding the optimal rate is NP-hard. Thus, most prior work focuses on obtaining bounds. 

The \textit{single uniprior} IC problem is a subclass of uniprior IC where the side-information sets at all receivers are singleton, and was studied in \cite{OnH}. In \cite{OnH}, a given single uniprior index coding problem is represented as a \textit{information flow graph}, containing a set of vertices representing each message (and thus the unique receiver which has it as side-information). The edges of the information flow graph represent the demands made by the receivers. The authors of \cite{OnH} characterized the length of an optimal index code, and also gave an optimal linear code construction for single uniprior IC. It was shown that the code can be constructed using a polynomial-time graphical algorithm. 

Unlike in the unicast case, a general uniprior problem cannot be always considered to be a single uniprior problem. In this work, we look at the general uniprior problem, where the side-information sets are disjoint but not necessarily singleton. In particular, our contributions are as follows.
\begin{itemize}
\item \textit{Demand Supergraphs: }We model the uniprior index coding problem using a \textit{demand supergraph}, consisting of supervertices and subvertices, and edges. Each supervertex represents a receiver and the subvertices of a supervertex are the set of side-information symbols available at that receiver. The edges represent the demands.
\item \textit{Bounds for Generalized Cycles: }We focus on uniprior IC problems on a special class of supergraphs (which we call \textit{generalized cycles}). We exploit the relationship of such problems with unicast IC and obtain lower and upper bounds on the optimal broadcast rate of index codes for such problems (Theorem \ref{upperlowerbounds}). 
\item \textit{NP-Hardness and Explicit code structures:} Bounds for generalized cycles: Using an equivalence between generalized cycles and Eulerian directed graphs, we show the class of generalized cycles for which the lower and upper bounds on the optimal rate are met with equality (Theorem \ref{GenCycleEulerianbounds}). We also show the explicit structure of a feasible scheme for such uniprior problems (Theorem \ref{explicitcode}). Further, we also show the NP-hardness of determining the lower bound using this equivalence. (Theorem \ref{corrtaue}). Thus, unlike single uniprior IC, obtaining optimal index codes for general uniprior IC or even characterising the optimal broadcast rate could be NP-hard.
\item \textit{Extending generalized cycles:} We then generalize the special class to a larger class of uniprior problems, obtain bounds on the optimal rate, and show a feasible index coding scheme for the larger class (Theorem \ref{thmgencycleextension}).
\end{itemize}
\textit{Notations and a few basic definitions: } A directed graph along with its vertex and edge sets is represented as ${\cal G}({\cal V},{\cal E}).$ A union of two graphs is a union of the set of vertices and the set of edges. A decomposition of a graph ${\cal G}$ is a set of subgraphs which are edge-disjoint, and whose union gives the graph ${\cal G}$. A trail of a directed graph ${\cal G}({\cal V},{\cal E})$ is a list of distinct edges $e_1,...,e_L$ such that the $head(e_i)=tail(e_{i+1}), 1\leq i\leq L-1$. A trail is closed if its start vertex and end vertex are the same. A directed graph is strongly connected if there is a path from any vertex to any other vertex. For a directed graph ${\cal G}$, let $\nu_e({\cal G})$ denote the maximum number of edge-disjoint cycles in ${\cal G}$. The quantity $\nu_v({\cal G})$ is the maximum number of vertex-disjoint cycles in ${\cal G}$. A \textit{feedback vertex set} of a directed graph ${\cal G}$ is a  set of vertices whose removal leads to an acyclic graph. Let the size of a minimal feedback vertex set be denoted by $\tau_v({\cal G})$. Similarly, the size of a minimal feedback edge set is denoted by $\tau_e({\cal G}$). For a graph with vertex set ${\cal V}$, we have that $\alpha({\cal G})=|{\cal V}|-\tau_v({\cal G})$ is the number of vertices in a maximum acyclic induced subgraph of ${\cal G}$. An underlying undirected graph of a given directed graph ${\cal G}$ is the undirected graph obtained by ignoring the directions in ${\cal G}$. An undirected graph $H$ is called a minor of an undirected graph $G$ if $H$ can be obtained from $G$ by a series of edge contractions and deletions. For more preliminaries on graphs, the reader is referred to \cite{DBW}. A finite field with $q$ elements is denoted by ${\mathbb F}_q$.

Due to space restrictions, some of the proofs have been omitted, but made available in \cite{MaK}.
\section{Preliminaries : Index Coding and Single Unicast}
\label{sec1}
Formally, the index coding (IC) problem (over some field $\mathbb{F}_q$) consists of a broadcast channel which can carry symbols from ${\mathbb F}_q$, along with the following.
\begin{itemize}
\item A set of $m$ receivers
\item A source which has messages ${\cal X}=\{x_i, i\in[1:n]\}$, each of which is modelled as a $t$-length vector over ${\mathbb F}_q$.
\item For each receiver $j$, a set $D(j)\subseteq {\cal X}$ denoting the set of messages demanded by the receiver $j$.
\item For each receiver $j$, a set $S(j)\subseteq {\cal X}\backslash D(j)$ denoting the set of $s_j$ side-information messages available at the $j^{th}$ receiver. 
\end{itemize}

For a message vector $\boldsymbol{x}\in{\mathbb F}^{nt}$, the source transmits a $l$-length codeword ${\mathbb E}(\boldsymbol{x})$ (the function $\mathbb E:{\mathbb F}^{nt}\rightarrow {\mathbb F}^l$, is known as the \textit{index code}), such that all the receivers can recover their demands. The quantity $l$ is known as the \textit{length of the code} $\mathbb E$. The \textit{transmission rate} of the code is defined as $\frac{l}{t}$. If $t=1$, then the index code is known as a \textit{scalar index code}, else it is known as a \textit{vector index code}. A linear encoding function ${\mathbb E}$ is also called a linear index code. The goal of index coding is to find optimal index codes, i.e., those with the minimum possible transmission rate. For an index coding problem ${\cal I}$ (over ${\mathbb F}_q$) with $t$-length messages, let $\beta_q(t,{\cal I})$ denote the length of an optimal vector index code. The \textit{broadcast rate} \cite{BKL2} is then $\beta_q({\cal I})=\lim_{t\rightarrow \infty}\frac{\beta_q(t,{\cal I})}{t}$. Clearly, we have $\beta_q({\cal I})\leq \beta_q(1,{\cal I}).$

An index coding problem is called a \textit{single unicast} problem if $m=n$ and each message is demanded by exactly one receiver. An index coding problem is called a general uniprior (or simply, a \textit{uniprior}) problem if $S(j)\cap S(j')=\phi, \forall j\neq j'.$ The single uniprior problem is then a special case of the uniprior problem with $s_j=1$. 

A given single unicast index coding problem ${\cal I}$ can be modelled using a directed graph called the \textit{side-information graph} \cite{BBJK}, denoted by ${\cal G}_{SI}({\cal V}_{SI},{\cal E}_{SI})$, where the set of vertices ${\cal V}_{SI}$, identified with the set of message symbols  ${\cal X}$, represents also the the set of receivers (each demanding an unique message). A directed edge  $(x_j,x_i)$ in ${\cal E}_{SI}$ indicates the availability of the message symbol $x_i$ as side-information at the receiver $j$ (which demands $x_j$). It was shown in \cite{BBJK} that the length of any optimal scalar linear index code (over ${\mathbb F}_q$) is equal to a property of the graph ${\cal G}_{SI}$ called the minrank, denoted by $mrk_q({\cal G}_{SI})$. While computing $mrk_q({\cal G}_{SI})$ is known to be NP-hard \cite{Pee} in general, several authors have given lower bounds and upper bounds for the quantity, as well as specific graph structures for which the bounds are met with equality (see for example, \cite{BBJK,DSC,BKL2}). From \cite{BBJK,DSC,BKL2}, we know that given a single unicast IC problem ${\cal I}$ on ${\cal G}_{SI}$ (with $n$ message vertices), we have the following.
\begin{align}
\label{eqn5}
n-\tau_v({\cal G}_{SI})\hspace{-0.05cm}\leq \hspace{-0.05cm}\beta_q({\cal I})\hspace{-0.05cm} \leq \hspace{-0.05cm}\beta_q(1,{\cal I})\hspace{-0.05cm} \leq mrk_q({\cal G}_{SI})\hspace{-0.05cm}\leq n-\nu_v({\cal G}_{SI}).
\end{align}
All the above quantities are NP-hard to compute for general graphs \cite{Pee,Kar}.

\section{Uniprior Index Coding: Modeling and Bounds}

\subsection{Modeling Uniprior IC using the Demand Supergraph}
\begin{definition}
For a given uniprior IC problem ${\cal I}$ with message set ${\cal X}$, we define a supergraph ${\cal G}_s({\cal V}_s,{\cal X},{\cal E}_s)$ as follows. 
\begin{itemize}
    \item For receiver $j$ in ${\cal I}$, there exists a corresponding supervertex $j\in  {\cal V}_s$.
    \item Each supervertex $j$ contains subvertices indexed by the side-information $S(j)\subset {\cal X}.$  
    \item An edge $(x_i,j)\in {\cal E}_s$ with tail node being the subvertex $x_i$ and head node being supervertex $j$ denotes that the message $x_i$ is demanded by the receiver $j$. All such demands in ${\cal I}$ are represented by their corresponding edges in the super graph.
\end{itemize}
\end{definition}
The notation $x^j$ denotes some arbitrary message (subvertex) in $S(j)$. For $x^j \in S(j),$ we also use $x^j\in j$ with respect to the supergraph. We also denote by $l^*({\cal G}_s)$  the length of an optimal scalar linear index code for the uniprior IC problem defined by ${\cal G}_s$, and use the term $l^*$ instead when there is no confusion. We now give the definition of a cycle in a supergraph.
\begin{definition}
A cycle (of length $L$) in a supergraph ${\cal G}_s$ is a sequence of distinct edges of ${\cal G}_s$ of the form  \begin{equation*}
\mathcal{C}=((x^{i_0},i_1),(x^{i_1},i_2),..,(x^{i_{L-1}},i_0))
\end{equation*}
where $x^{i_{j}} \in i_{j}, \forall j=0,...,L-1$ and the supervertices $i_j, 0\leq j\leq L-1$ are all distinct. 
A supergraph without a cycle is naturally called an \textit{acyclic} supergraph.
\end{definition}
\subsection{A special uniprior problem which is also single unicast}
We now define a class of uniprior IC problems which are also a special case of single unicast IC problems.
\begin{definition}[Generalized Cycle] 
\label{gencycle}
A  \textit{generalized cycle} denoted by ${\cal G}_{gc}({\cal V}_{gc},{\cal X}_{gc},{\cal E}_{gc})$ is a demand supergraph satisfying the following properties. 
\begin{itemize}
    \item The message set is ${\cal X}_{gc}$ and each message (subvertex) is demanded exactly once.
    \item The number of incoming edges to any supervertex $j \in {\cal V}_{gc}$ is equal to $s_j$ (the number of side-information symbols).
    \item The supervertices ${\cal V}_{gc}$ are connected, i.e., for every two $i,j\in {\cal V}_{gc}$, there is a path from some message in $i$ to $j$.
\end{itemize}
\end{definition}
\begin{remark}
For a single uniprior index coding problem, the definition of a demand supergraph specializes to the information flow graph of \cite{OnH}, and the definition of a generalized cycle specializes to a cycle in the information flow graph.
\end{remark}
Given a uniprior IC problem ${\cal I}$ on a generalized cycle, it is clear that ${\cal I}$ can be looked at as a single unicast problem also (making $s_j$ `copies' of a receiver $j$, each demanding an unique single symbol in $D(j)$).  Hence one can define its corresponding side-information graph ${\cal  G}_{SI}$. It is easy to see that for each message $x_i$ demanded by a receiver $j$ in ${\cal G}_{gc}$, there exists $s_j$ edges in the corresponding ${\cal G}_{SI}$ to each message (subvertex) in $S(j)$ from $x_i$.

Equivalent to definition of $\nu_e({\cal G})$ for a directed graph ${\cal G}$, let $\nu_e({\cal G}_s)$ be the maximum number of edge-disjoint cycles of ${\cal G}_s$. We now prove a result which shows that edge-disjoint cycles of ${\cal G}_{gc}$ are equivalent to vertex-disjoint cycles of the corresponding ${\cal G}_{SI}$ and vice-versa. 
\begin{proposition}
\label{cktsGCSI}
Consider an uniprior IC problem with its demand supergraph being a generalized cycle ${\cal G}_{gc}$, and the corresponding side-information graph ${\cal G}_{SI}$. For any set ${\cal C}$ of edge-disjoint cycles in ${\cal G}_{gc}$, there exist a set of ${\cal C}'$ vertex-disjoint cycles in ${\cal G}_{SI}$ of the same cardinality, and vice versa. Thus $\nu_e({\cal G}_{gc})=\nu_v({\cal G}_{SI})$.
\end{proposition}
\begin{IEEEproof}
We prove the theorem for a set of two cycles. The extension to any finite number of cycles follows.

Suppose $\mathcal{C}_1, \mathcal{C}_2$ are any two edge-disjoint cycles in ${\cal G}_{gc}$, where 
\begin{align}
\label{eqn1}
\mathcal{C}_1=((x^{i_0},i_1), (x^{i_1},i_2),...,(x^{i_{L-1}},i_0)),\\ 
\label{eqn2}
\mathcal{C}_2=((x^{j_0},j_1), (x^{j_1},j_2),...,(x^{j_{L^{'}-1}},j_0)),   
\end{align}
where $x^{i_k}\in i_k, \forall k$ and $x^{j_{k_1}}\in j_{k_1}, \forall k_1$.

As the cycles are edge-disjoint, we must have $(x^{i_k},i_{(k+1)(mod~L)})\neq(x^{j_{k_{1}}},j_{(k_1+1) (mod~L^{'})}), \forall k,k_1$. Note that this implies $x^{i_k}\neq x^{j_{k_{1}}}$ for any $k,{k_1}$, as each message is demanded precisely once in ${\cal G}_{gc}$.

Consider the cycles correspondingly in ${\cal G}_{SI}$ considered as follows. 
\begin{align}
\label{eqn3}
\mathcal{C}_1'=((x^{i_0},x^{i_1}), (x^{i_1},x^{i_2}),...,(x^{i_{L-1}},x^{i_0})).\\ 
\label{eqn4}
\mathcal{C}_2'=((x^{j_0},x^{j_1}), (x^{j_1},x^{j_2}),...,(x^{j_{L^{'}-1}},x^{j_0})).   
\end{align}

Such cycles clearly exist because $x^{i_k}\in i_k, \forall k$ and $x^{j_{k_1}}\in j_{k_1}, \forall k_1.$ As $x^{i_k}\neq x^{j_{k_{1}}}$ for any $k,{k_1}$, it is clear that the cycles ${\cal C}_1'$ and ${\cal C}_2'$ in ${\cal G}_{SI}$ are vertex-disjoint.

The converse follows by picking vertex-disjoint cycles in ${\cal G}_{SI}$ as in (\ref{eqn3}) and (\ref{eqn4}) and showing that corresponding edge-disjoint cycles exist in ${\cal G}_{gc}$ as in (\ref{eqn1}) and (\ref{eqn2}). We leave the details to the reader.
\end{IEEEproof}
\begin{remark}
It is easy to see that Proposition \ref{cktsGCSI} should hold for all uniprior IC problems whose supergraph satisfies the first property of Definition \ref{gencycle}. For the purposes of this work, Proposition \ref{cktsGCSI} is sufficient.
\end{remark}
\begin{figure}[ht]
\centering
\includegraphics[height=2.5in]{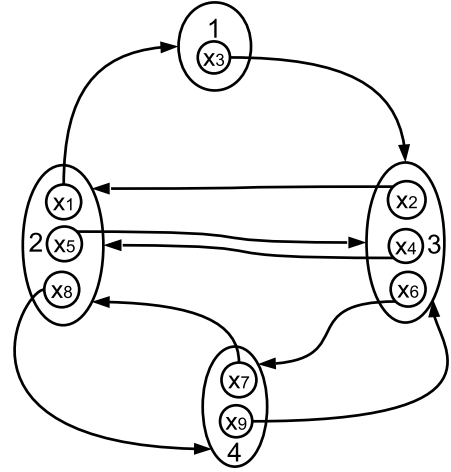}
\caption{Supergraph corresponding to uniprior IC problem of Example \ref{exm1}}
\label{fig:supergraph}
\end{figure}
\begin{figure}[ht]
\centering
\includegraphics[height=2.2in]{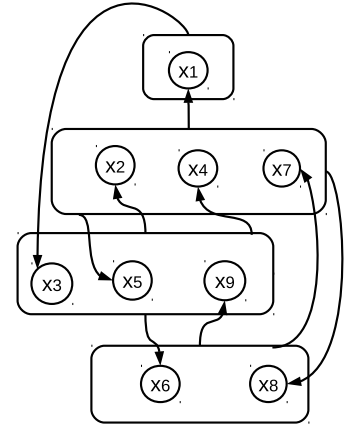}
\caption{Side-information graph for the uniprior IC problem of Fig. \ref{fig:supergraph}. An edge incoming at any $x_i$ and from some `block' indicates that there are edges incoming at $x_i$ from each message in the block}
\label{fig:sideinfograph}
\end{figure}
\begin{example}
\label{exm1}
Consider a uniprior index coding problem with nine messages and four receivers. The demand sets, side information sets are as follows: $D(1)=x_1,$ $S(1)=x_3$, $D(2)=\{x_2,x_4,x_7\},$  $S(2)=\{x_1,x_5,x_8\},$ $D(3)=\{x_3,x_5,x_9\},$ $S(3)=\{x_2,x_4,x_6\}$, $D(4)=\{x_6,x_8\}$, $S(4)=\{x_7,x_9\}$. The demand super graph corresponding to the problem is a generalized cycle, shown in Fig. \ref{fig:supergraph}. The demand supergraph ${\cal G}_{gc}$ in Fig. \ref{fig:supergraph} can be decomposed in to four edge-disjoint cycles : $\big((x_1,1),(x_3,3),(x_2,2)\big)$, $\big((x_5,3),(x_4,2))\big)$, $\big((x_9,3),(x_6,4)\big),$ and $\big((x_7,2),(x_8,4)\big)$, denoted by ${\cal C}_1$, ${\cal C}_2,$ ${\cal C}_3$ and ${\cal C}_4$ respectively. 
Corresponding to the cycles ${\cal C}_i$, we get the cycles $\big((x_1,x_3),(x_3,x_2),(x_2,x_1)\big)$, $\big((x_5,x_4),(x_4,x_5)\big)$, $\big((x_9,x_6),(x_6,x_9)\big)$, and $\big((x_7,x_8),(x_8,x_7)\big)$ in the side-information graph shown in Fig. \ref{fig:sideinfograph}. 
\end{example}

Following the definition of a feedback vertex set of a directed graph, we define the feedback edge set of a demand supergraph ${\cal G}_s$ as a set of edges whose removal leads to an acyclic supergraph. The size of a minimal feedback edge set is denoted by $\tau_e({\cal G}_s)$. 
\begin{proposition}
\label{MFVequalsMFEthm}
Let a generalized cycle ${\cal G}_{gc}$ represent a uniprior IC problem ${\cal I}$, and let ${\cal G}_{SI}$ be its corresponding side-information graph. Suppose the set of edges \begin{equation}
\label{fbedgeset}
    \{(x^{i_k},j_k):k=1,..,K, j_k\in {\cal V}_{gc}\}
\end{equation}
is a feedback edge set of ${\cal G}_{gc}$. Then the set of vertices
\begin{equation}
\label{fbvertexset}
\{x^{i_k}:k=1,..,K\}
\end{equation} 
is a feedback vertex set of ${\cal G}_{SI}$. Conversely, if (\ref{fbvertexset}) is a feedback vertex set of ${\cal G}_{SI}$, then for some $K$ supervertices $\{j_k:k=1,..,K\}$ in ${\cal G}_{gc}$, the set in (\ref{fbedgeset}) is a feedback edge set of ${\cal G}_{gc}$. Thus, $\tau_v({\cal G}_{SI})=\tau_e({\cal G}_{gc})$.
\end{proposition}
\begin{IEEEproof}
We first show that if (\ref{fbedgeset}) is a feedback edge set of ${\cal G}_{gc}$, then (\ref{fbvertexset}) must be a feedback vertex set of ${\cal G}_{SI}$. Let ${\cal G}_{SI}'$ denote the subgraph of ${\cal G}_{SI}$ which remains after deleting the vertices $\{x^{i_k}:k=1,..,K\}$ (and the incident edges on them). Suppose ${\cal G}_{SI}'$ is not acyclic, then there exists a cycle $\cal C^{'}$ in ${\cal G}_{SI}$ that does not have any vertices from $\{x^{i_k}:k=1,..,K\}$. Then by Proposition \ref{cktsGCSI}, there exists a corresponding cycle $\cal C$ in ${\cal G}_{gc}$ which does not have the edges in (\ref{fbedgeset}). This means (\ref{fbedgeset}) is not a feedback edge set of ${\cal G}_{gc}$, which is a contradiction. The converse can be similarly proved; we leave this to the reader.
\end{IEEEproof}
We now give the main theorem in this section. 
\begin{theorem}
\label{upperlowerbounds}
For an uniprior IC problem ${\cal I}$ on a generalized cycle ${\cal G}_{gc}$, we have
\begin{align}
\label{boundforgeneralizedcycles}
n-\tau_e({\cal G}_{gc})\hspace{-0.05cm}\leq \hspace{-0.05cm}\beta_q({\cal I})\leq \hspace{-0.05cm}\beta_q(1,{\cal I})\hspace{-0.05cm}\leq l^*\hspace{-0.05cm}\leq n-\nu_e({\cal G}_{gc}).
\end{align}
\end{theorem}
\begin{IEEEproof}
As ${\cal I}$ is also a single unicast problem (represented by, say, ${\cal G}_{SI}$) and by Proposition \ref{cktsGCSI} and Proposition \ref{MFVequalsMFEthm}, we have   
\begin{align*}
\nu_v({\cal G}_{SI})&=\nu_e({\cal G}_{gc}),\\
\tau_v({\cal G}_{SI})&=\tau_e({\cal G}_{gc}).
\end{align*}
Furthermore, $l^*=mrk_q({\cal G}_{SI})$. With all these facts, we can invoke (\ref{eqn5}) to prove our theorem.
\end{IEEEproof}
\section{Explicit codes, Hardness results, and Tightness of Bounds}
In this section, we show an explicit achievable index code, and also show the NP-hardness of obtaining the lower bound in Theorem \ref{upperlowerbounds}, and obtain a special class of generalized cycles for which (\ref{boundforgeneralizedcycles}) is satisfied with equality throughout. For this purpose we use the connection between Eulerian directed graphs and generalized cycles.

\subsection{Eulerian Directed Graphs}
Eulerian graphs \cite{DBW} are those which contain an Eulerian circuit, which is a closed trail containing all edges. The following Lemma is found in \cite{DBW} (Chapter 1), and will be used in this section.%
\begin{lemma}
\label{eulerianincoutgequal}
A directed graph (with at least one edge incident on each vertex) is Eulerian if and only if for every vertex, the number of incoming edges is equal to the number of outgoing edges and the graph is strongly connected.
\end{lemma}

The first statement of the following lemma is also known as Veblen's theorem for directed graphs (\cite {DBW}, Chapter 1, Exercise 1.4.5). The second statement is mentioned in \cite{Sey} in passing. As a formal statement or proof could not be found, we include a short proof here.
\begin{lemma}
\label{euleriandecomposition}
An Eulerian directed graph ${\cal G}$ can be decomposed into a set of edge-disjoint cycles. In particular, any maximal set of edge-disjoint cycles of ${\cal G}$ is also a decomposition of ${\cal G}$.
\end{lemma}
\begin{IEEEproof}
We only prove the second statement as it implies the first statement. Suppose some maximal set $\mathfrak C$ of edge-disjoint cycles is not a decomposition. Assume that we remove all the edges from ${\cal G}$ which are present in $\mathfrak C$, and subsequently also any isolated vertices (which don't have incoming or outgoing edges after the removal of $\mathfrak C$) and look at the remaining graph ${\cal G}'$. By our assumption that $\mathfrak C$ is not a decomposition, ${\cal G}'$ must have least one edge and thus at least two vertices. However also note that, for any remaining vertex, the number of incoming and outgoing edges must be the same (since  for any removed incoming edge, one outgoing edge must also be removed). This means that ${\cal G}'$ is also Eulerian, which means there is one cycle in ${\cal G}'$ (and hence in ${\cal G}$) which is edge-disjoint to those in $\mathfrak C$. This contradicts the maximality of $\mathfrak C$ and concludes the proof.
\end{IEEEproof}
\subsection{Eulerian Graphs associated with Generalized Cycles}
\begin{definition}[Eulerian Graph associated with ${\cal G}_{gc}$]
The Eulerian graph associated with a generalized cycle ${\cal G}_{gc}({\cal V}_{gc},{\cal X}_{gc},{\cal E}_{gc})$ is the directed graph ${\cal G}_{eu}$ with vertex set ${\cal V}_{eu}={\cal V}_{gc}$ and edge set ${\cal E}_{eu}$ defined as follows.
\begin{itemize}
\item For each edge $(x^{i},j)$ with message $x^{i}\in i$, an edge from $i$ to $j$ exists in ${\cal G}_{eu}.$ 
\end{itemize}
\end{definition}
\begin{remark}
It is easy to see that the directed graph defined in the above way is indeed Eulerian, by the properties of ${\cal G}_{gc}$ (using Lemma \ref{eulerianincoutgequal}). From the definition it should be clear that the Eulerian graph associated with a generalized cycle could have parallel edges. Suppose there exists $p$ edges between the vertices $i$ and $j$ in the Eulerian graph ${\cal G}_{eu}$, then we refer to those edges as $\{(i,j)_k : k=1,2,...,p\}$.
\end{remark}
\begin{figure}[htbp]
\centering
\includegraphics[height=1.5in]{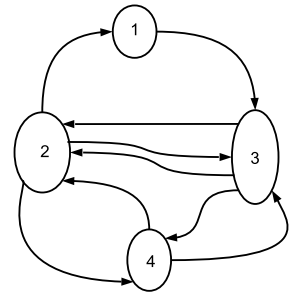}
\caption{Eulerian Graph associated with the generalized cycle in Fig. \ref{fig:supergraph}}
\label{fig:eulerianexamplefig}
\end{figure}
Fig. \ref{fig:eulerianexamplefig} represents the Eulerian graph associated with the generalized cycle in Fig. \ref{fig:supergraph}.
We can thus obtain a Eulerian graph from any given generalized cycle. The following lemma shows that there is a generalized cycle corresponding to each Eulerian graph. 
\begin{lemma}
\label{lemmaEulerianimpliesGgc}
Let ${\cal G}$ be an Eulerian directed graph with no isolated vertices (each vertex has at least one incident edge). Then there exists a generalized cycle ${\cal G}_{gc}$ whose equivalent Eulerian graph is ${\cal G}$.
\end{lemma}
\begin{IEEEproof}
Let ${\cal G}(\cal V, \cal E)$ be the given Eulerian graph. We construct a corresponding generalized cycle ${\cal G}_{gc}$ with ${{\cal V}_{gc}}={\cal V}$ as follows. Assume that the sets ${\cal V}_{gc}$ and ${\cal V}$ are indexed from $1$ to $m=|{\cal V}|$.

Consider a vertex $i\in {\cal G}$ having $p$ outgoing edges equal (and thus $p$ incoming edges as well). In the corresponding supervertex $i\in {\cal G}_{gc}$, we create $p$ subvertices. For an edge $(i,j)$ in ${\cal G}$, we construct an edge in ${\cal G}_{gc}$ starting from a unique message subvertex of supervertex $i$ from which no previous outgoing edge exists, ending at supervertex $j$. Clearly, the number of incoming and outgoing edges at any supervertex $i$ of ${\cal G}_{gc}$ is the same as that of $i\in{\cal V}$. Furthermore, as ${\cal G}$ is Eulerian and has no isolated vertices, it is strongly connected. Thus the ${\cal G}_{gc}$ so constructed is a generalized cycle.
\end{IEEEproof}

\subsection{Using Eulerian graphs to show a simple explicit code}
The following proposition relates some properties of ${\cal G}_{gc}$ with ${\cal G}_{eu}$. 
\begin{proposition}
\label{thmgencycleEQUIVeulerian}
Let ${\cal G}_{gc}$ be a generalized cycle and ${\cal G}_{eu}$ be the corresponding Eulerian graph. For any set ${\cal C}$ of edge-disjoint cycles in ${\cal G}_{gc}$, there exist a set ${\cal C}'$ of edge-disjoint cycles in ${\cal G}_{eu}$ of the same cardinality, and vice versa. Thus 
\begin{align}
\label{nuGgcGeu}
\nu_e({\cal G}_{gc})=\nu_e({\cal G}_{eu}).
\end{align}
Furthermore, we also have 
\begin{align}
\label{tauGgcGeu}
\tau_e({\cal G}_{gc})=\tau_e({\cal G}_{eu}).
\end{align}
\end{proposition} 
%
\begin{IEEEproof}
We proceed in a similar way as in the proof of Proposition \ref{cktsGCSI}, using the properties of the generalized cycle to prove our result. We consider two edge-disjoint cycles in ${\cal G}_{gc}$ as in (\ref{eqn1}) and (\ref{eqn2}). 
We claim that corresponding edge-disjoint cycles can be picked in ${\cal G}_{eu}$ as follows. 
\begin{align}
\label{eqn7}
\mathcal{C}^{'}_{1}=((i_0,i_1)_{m_0},(i_1,i_2)_{m_1},...(i_{L-1},i_0)_{m_{L-1}})\\
\label{eqn8}
\mathcal{C}^{'}_{2}=((j_0,j_1)_{n_0},(j_1,j_2)_{n_1},...,(j_{L^{'}-1},j_0)_{n_{L^{'}-1}})  
\end{align}
where $(i_k,i_{(k+1)(mod~L)})_{m_k}$ is the ${m_k}^{th}$ edge between $i_k$ and $i_{(k+1)(mod~L)}$ in ${\cal G}_{eu}$, and similarly  $(j_{k^{'}},j_{(k^{'}+1)(mod~L^{'})})_{n_{k^{'}}}$ is the  ${n_{k^{'}}}^{th}$ edge between vertices  $j_{k^{'}}$ and $j_{(k^{'}+1)(mod~L^{'})}.$ 

To see why such edge-disjoint cycles can be picked, we first note that an edge can be common between the two cycles ${\cal C}_1'$ and ${\cal C}_2'$ only when $i_k=j_{k'}$ and $i_{(k+1)(mod~L)}=j_{(k'+1)(mod~L)}$, for some $(x^{i_k},i_{(k+1)(mod~L)})\in {\cal C}_1$ (with $x^{i_k}\in i_k$), and $(x^{j_{k'}},j_{(k'+1)(mod~L)})\in {\cal C}_2$ (with $x^{j_{k'}}\in j_{k'}$). However, in such a scenario, there must be two distinct edges in ${\cal G}_{eu}$ between $i_k$ and $i_{(k+1)(mod~L)}$, because the two edges in ${\cal C}_1$ and ${\cal C}_2$ are distinct. We can therefore find $m_k, n_{k'}$ such that $m_k\neq n_{k'}$, thus getting $(i_k,i_{(k+1)(mod~L)})_{m_k}\neq (j_{k^{'}},j_{k^{'}+1 (mod~L^{'})})_{n_{k^{'}}}$. Continuing this way, we can choose two edge-disjoint cycles ${\cal C}_1$ and ${\cal C}_2$.

%
Conversely consider two edge-disjoint cycles from ${\cal G}_{eu}$ as in (\ref{eqn7}) and (\ref{eqn8}).
Corresponding to cycle ${\cal C}'_1$ in ${\cal G}_{eu}$, it is easy to see that there is a cycle ${\cal C}_1$ in ${\cal G}_{gc}$ as follows. 
\begin{equation*}
\mathcal{C}_1=((x^{i_0,1},i_1), (x^{i_1,1},i_2),...,(x^{i_{L-1},1},i_0))
\end{equation*}
where $x^{i_{k},1}\in i_k$ is some message symbol available at supervertex $i_k$ and demanded by $i_{(k+1)(mod~L)}$. The vertices $x^{i_{k},1}, \forall k$ are well-defined because if an edge $(i_k,i_{(k+1)(mod~L)})$ exists in ${\cal G}_{eu}$, then at least one message $x^{i_{k},1}\in i_k$ must exist and be demanded by $i_{(k+1)(mod~L)}$.

Now with respect to the cycle ${\cal C}_2'$, we pick a cycle ${\cal C}_2$ in ${\cal G}_{eu}$ which is edge-disjoint from ${\cal G}_1$ as follows. For some edge $(j_{k'},j_{(k'+1)(mod~L')})\in {\cal C}_2'$, if there is no $(i_k,i_{(k+1)(mod~L)})\in {\cal C}_1'$ such that $j_{k'}=i_k$, then we pick an edge $(x^{j_{k'},2},j_{(k'+1)(mod~L')})$ for some message $x^{j_{k'},2}\in j_{k'}$ which is demanded by $j_{(k'+1)(mod~L')}$ (such a message indeed exists). 

On the other hand, if there exists an $(i_k,i_{(k+1)(mod~L)})\in {\cal C}_1'$ such that $i_k=j_{k'}$, then because the cycles ${\cal C}_1'$ and ${\cal C}_2'$ are edge-disjoint, there must be a message $x^{j_{k'},2}\in j_{k'}$ which is distinct from $x^{i_{k},1}$ and is demanded by $j_{(k'+1)(mod~L')}$. We thus choose the edge $(x^{j_{k'},2},j_{(k'+1)(mod~L')})$. This way, we can construct ${\cal C}_2$, corresponding to ${\cal C}_2'$, and edge-disjoint of ${\cal C}_1$ in ${\cal G}_{gc}$. This completes the proof of (\ref{nuGgcGeu}). The proof of (\ref{tauGgcGeu}) is similar as that of Proposition \ref{MFVequalsMFEthm}, hence we skip it.

\end{IEEEproof}
Theorem \ref{upperlowerbounds} gave an upper bound for the generalized cycle IC problem and it is clear that an achievable scheme based on the circuit packing bound on ${\cal G}_{SI}$ of \cite{DSC} meets the upper bound. The following theorem makes the structure of an achievable code explicit. For this, we need the idea of encoding messages in a cycle, which is well known in index coding literature for the unicast problem. For a cycle $\mathcal{C}=((x^{i_0},i_1),(x^{i_1},i_2),..,(x^{i_{L-1}},i_0))$ of ${\cal G}_{gc}$, we refer to the set of $L-1$ transmissions $x^{i_0}-x^{i_1},x^{i_1}-x^{i_2},...,x^{i_{L-2}}-x^{i_{L-1}},$ as the \textit{cyclic code} associated with ${\cal C}$. 
\begin{theorem}
\label{explicitcode}
Let ${\cal I}$ be an uniprior IC problem on a generalized cycle ${\cal G}_{gc}$, and $\nu=\nu_e({\cal G}_{gc})$. Let ${\mathfrak C}=\{{\cal C}_1,{\cal C}_2,...,{\cal C}_{\nu}\}$ be a maximal set of edge-disjoint cycles of ${\cal G}_{gc}$. The transmissions corresponding to the cyclic code associated with each of the edge-disjoint cycles in ${\mathfrak C}$ is an index code for ${\cal I}$ with length $n-\nu$.
\end{theorem}
\begin{IEEEproof}
By Lemma \ref{euleriandecomposition} and by Proposition \ref{thmgencycleEQUIVeulerian}, the set of cycles in ${\mathfrak C}$ decompose ${\cal G}_{gc}$, in the sense that all the edges (and hence all the messages as well) of ${\cal G}_{gc}$ must lie in the cycles in $\mathfrak C$. Consider the cyclic code associated with any cycle ${\cal C}\in {\mathfrak C}$. This code satisfies all the demands corresponding to the edges in $\cal C$. Thus the cyclic codes associated with all the cycles in $\mathfrak C$ satisfy all the demands in ${\cal G}_{gc}$. The number of transmissions is $n-\nu$ (as one transmission is `saved' for each cycle, and the cycles in $\mathfrak C$ are edge-disjoint). This proves the theorem.
\end{IEEEproof}
\begin{remark}
We call the code as described by Theorem \ref{explicitcode} as the \textit{cyclic code} corresponding to the generalized cycle ${\cal G}_{gc}$.
\end{remark}

\subsection{Tightness of the bounds of Theorem \ref{upperlowerbounds} and NP-hardness of the lower bound}
If ${\cal G}_{gc}$ consists of just a single cycle, then it is clear that the lower and upper bounds of Proposition \ref{MFVequalsMFEthm} coincide. Since the generalized cycle ${\cal G}_{gc}$ is highly structured and has properties similar to cycles, it may be tempting to think that  the same holds for all generalized cycles. The following proposition from \cite{Sey} will be used to show that there exists generalized cycles for which the upper and lower bound shown in Theorem \ref{upperlowerbounds} are not always equal. We note before that the Petersen family of graphs (shown in \cite{Sey}) are a set of seven graphs which can be obtained by transformations of the Petersen graph. 
\begin{proposition}\cite{Sey} 
\label{petersoneulerian}
Let ${\cal G}$ be an Eulerian directed graph, such that its underlying undirected graph has no minor in the Petersen family. Then $\nu_e({\cal G})=\tau_{e}({\cal G})$. Also, there exists Eulerian graphs for which $\nu_e({\cal G})<\tau_{e}({\cal G})$. 
\end{proposition}
We thus get the following result as a direct consequence of Proposition \ref{petersoneulerian}, Proposition \ref{thmgencycleEQUIVeulerian} and Lemma \ref{lemmaEulerianimpliesGgc}.
\begin{theorem}
\label{GenCycleEulerianbounds}
Let the generalized cycle ${\cal G}_{gc}$ represent a uniprior IC problem ${\cal I}$ with $n$ messages. Suppose the underlying undirected graph of the equivalent Eulerian graph ${\cal G}_{eu}$ of ${\cal G}_{gc}$ does not have a minor in the Petersen family. Then, we have 
\[
n-\tau_e({\cal G}_{gc})=\beta_q({\cal I})=\beta_q(1,{\cal I})=l^*=n-\nu_e({\cal G}_{gc}),
\]
for any field size $q$. There also exist uniprior IC problems on generalized cycles for which $\nu_e({\cal G}_{gc})<\tau_{e}({\cal G}_{gc})$, and hence (\ref{boundforgeneralizedcycles}) is not satisfied with equality throughout.
\end{theorem}
\begin{example}
The generalized cycle ${\cal G}_{gc}$ in Fig. \ref{fig:supergraph} has the the corresponding Eulerian graph ${\cal G}_{eu}$ in Fig. \ref{fig:eulerianexamplefig} which does not have a minor in the Petersen family; and hence $\nu_e({\cal G}_{gc})=\tau_e({\cal G}_{eu})=4$ (we leave it to the reader to check these statements). Thus, the cyclic code corresponding to the cycles picked in Example \ref{exm1} is an optimal code with length $5$.  
\end{example}
As the final result in this section, we show hardness of finding $\tau_e({\cal G}_{gc})$. In a recent work \cite{PeP}, it was shown that finding $\tau_e({\cal G})$ for a directed Eulerian graph ${\cal G}$ is NP-hard. Thus we have the following result, by invoking Proposition \ref{thmgencycleEQUIVeulerian}. 
\begin{theorem}
\label{corrtaue}
Finding $\tau_e({\cal G}_{gc})$ is NP-hard. 
\end{theorem}
We also remark that finding $\nu_e({\cal G})$ for any graph (directed or undirected) is generally NP-hard (see \cite{KNSYY}, for example), and to the best of our knowledge there are no specific results about the complexity of finding $\nu_e({\cal G})$ for Eulerian directed graphs.
\section{Beyond generalized cycles}
From our results in the previous sections, we have developed a framework for studying the uniprior IC problem with the basic component being a generalized cycle. As a first step towards enlarging our understanding of the uniprior class of problems, we present a simple extension of the generalized cycle. Before the definition, we note that the subgraph of a demand supergraph is naturally defined as consisting of a subset of supervertices, subset of subvertices, and a subset of edges that run between them.
\begin{definition}
\label{decomposable}
Let ${\cal G}_s$ be a supergraph, such that ${\cal G}_s$ contains as a subgraph a generalized cycle ${\cal G}_{gc}$ containing all the vertices of ${\cal G}_s$. Let $\mathfrak C$ be a maximal set of edge-disjoint cycles of ${\cal G}_{gc}$. Then ${\cal G}_s$ is said to be demand-decomposable by $\mathfrak C$ if the following condition is satisfied.
\begin{itemize}
\item Any edge, which is present in ${\cal G}_s$ but not in any of the cycles of $\mathfrak C$, starts from some message subvertex in some cycle ${\cal C}\in{\mathfrak C}$ and ends at some supervertex in the same cycle ${\cal C}$.
\end{itemize}
\end{definition}
Thus, Definition \ref{decomposable} means that ${\cal G}_s$ is demand-decomposable by $\mathfrak C$ if no edges exist `across' different cycles in $\mathfrak C$. We now give the main result of this section. 
\begin{theorem}
\label{thmgencycleextension}
Consider a uniprior IC problem ${\cal I}$ (with $n$ messages) represented by a supergraph ${\cal G}_s$. Let ${\cal G}_{gc}$ and $\mathfrak C$ be as in Definition \ref{decomposable}. If ${\cal G}_s$ is demand-decomposable by $\mathfrak C$, then we have (for any field size $q$)
\begin{align}
\label{bounddecomposable}
n-\tau_e({\cal G}_{gc})\leq \beta_q({\cal I})\leq \beta_q(1,{\cal I})\leq l^*\leq n-\nu_e({\cal G}_{gc}).
\end{align}
Furthermore, (\ref{bounddecomposable}) is satisfied with equality throughout if the the underlying undirected graph of the Eulerian graph corresponding to ${\cal G}_{gc}$ has no minor in the Petersen family.
\end{theorem}
\begin{IEEEproof}
We prove the lower bound first. Note that removing edges (i.e. demands) from ${\cal G}_{s}$ cannot increase the optimal broadcast rate. Assume that we remove all edges from ${\cal G}_{s}$ which are not present in ${\cal G}_{gc}$, and let ${\cal I}'$ be the uniprior IC problem corresponding to ${\cal G}_{gc}$. Clearly, $\beta_q({\cal I}')\leq \beta_q({\cal I}).$ Note that the number of messages in ${\cal G}_{gc}$ is still $n$. By Theorem \ref{upperlowerbounds}, we have $n-\tau_e({\cal G}_{gc})\leq \beta_q({\cal I}').$ This gives the lower bound.

Now the upper bound. Note that cyclic code for ${\cal G}_{gc}$ corresponding to $\mathfrak C$ is such that every supervertex (i.e. receiver) in any cycle $\cal C$ in $\mathfrak C$ can decode all the messages in $\cal C$. Since the start-message and the end-supervertex of any additional edge in ${\cal G}_{s}$ (and not in ${\cal G}_{gc}$) are both in an identical cycle, this means that all those demands (denoted by the additional edges) will also be satisfied. Hence the cyclic code scheme of Theorem \ref{explicitcode} is an achievable scheme for ${\cal G}_s$ also. This gives us the upper bound and thus (\ref{bounddecomposable}). The last claim follows by Theorem \ref{GenCycleEulerianbounds} and (\ref{bounddecomposable}). 
\end{IEEEproof}
\begin{example}
Consider an uniprior extension of the uniprior IC problem of Example \ref{exm1} by changing only the demands of receivers $1,2$ and $3$ as $D(3)=\{x_1,x_3,x_5,x_9\}$, $D(2)=\{x_3,x_2,x_4,x_7\}$, $D(1)=\{x_1,x_2\}$. Then the supergraph corresponding to this problem is demand-decomposable by the set of four cycles described in Example \ref{exm1}. Hence the cyclic code which is optimal for the IC problem in Example \ref{exm1} is optimal for this extension also.
\end{example}

\end{document}